\documentclass[superscriptaddress,groupedaddress,nofootnoteinbib,11pt]{article}
\pdfoutput=1 
\usepackage{jheppub}
\usepackage{orcidlink}

\makeatletter
\gdef\@fpheader{}
\makeatother

\usepackage{CJKutf8} 
\usepackage{mathtools,slashed,mathrsfs}
\usepackage[caption=false]{subfig}
\usepackage{dcolumn}
\usepackage{multirow}
\usepackage{tabularx}
\usepackage{booktabs}
\usepackage{bm}
\usepackage{amsmath}
\usepackage{comment}
\usepackage{setspace}
\usepackage[normalem]{ulem} 
\usepackage{enumerate}
\usepackage{siunitx}
\usepackage{float}
\usepackage{hyperref}

\begin{document}

\title{Vector Portals at Future Lepton Colliders}

\date{\today}
\author[a]{Sagar Airen \orcidlink{0000-0002-3646-6537},}
\author[a]{Edward Broadberry \orcidlink{0000-0003-0652-1862},}
\author[b]{Gustavo Marques-Tavares \orcidlink{0000-0002-1861-7936},}
\author[a]{Lorenzo Ricci \orcidlink{0000-0001-8704-3545}}

\affiliation[a]{Maryland Center for Fundamental Physics, University of Maryland, College Park, MD 20742}
\affiliation[b]{Department of Physics and Astronomy, University of Utah, Salt Lake City, UT 84112, U.S.A.}


\abstract{
We assess the sensitivity of future lepton colliders to weakly coupled vector dark portals (aka ``\( Z' \) bosons'') with masses ranging from tens of GeV to a few TeV. Our analysis focuses on dark photons and \( L_{\mu} - L_{\tau} \) gauge bosons. We consider both visible and invisible decay channels. We demonstrate that both high energy \(\mu\) colliders and future \( e^+e^- \) colliders, using the FCC-ee \(Z\)-pole and \(ZH\) operation modes as a benchmark, offer significant improvements in sensitivity. We find that both colliders can enhance the sensitivity to \( L_{\mu} - L_{\tau} \) bosons (for both visible and invisible decays) and to invisibly decaying dark photons by 1--2 orders of magnitude across the relevant mass range. Furthermore, we study the impact of forward \( \mu \) detectors at the \( \mu \)-collider on the sensitivity to both models.
}
\maketitle

\section{Introduction}

Planning an experimental program to succeed the LHC is likely to become the most crucial consideration over the next decade in particle physics. 
The front runner proposals for the next collider are lepton colliders, either electron-positron ($e^+e^-$) machines (such as the FCC-ee\cite{FCC:2018evy}, CEPC\cite{CEPCStudyGroup:2018ghi}, ILC~\cite{Behnke:2013xla}, CCC \cite{Vernieri:2022fae}, and CLIC \cite{Roloff:2018dqu}), or a $\mu^+ \mu^-$ collider \cite{Accettura:2023ked}, which has received a lot of recent attention. Future $e^+e^-$ colliders will run at different center of mass (CoM) energies, including Tera-Z and Mega-$H$ factories, and in some proposals (such as the FCC-ee and CEPC) are to be followed by a high-energy $pp$ collider targeting $100$ TeV CoM energy \cite{FCC:2018vvp, CEPC-SPPCStudyGroup:2015csa}. On the other hand, a high-energy muon collider is an ambitious proposal for a high-luminosity $\mu^+ \mu^-$ collider ($\mu$C) with CoM energy around $10$ TeV. Both of these proposals have a strong physics case that has been extensively investigated \cite{CEPC-SPPCStudyGroup:2015csa,FCC:2018byv,FCC:2018vvp,Accettura:2023ked,Roloff:2018dqu}.

The next generation of colliders will also offer exciting opportunities to improve on the intensity frontier, and can play an important role in looking for new physics near or below the weak scale. An important target in this energy range are dark sectors, which are well motivated candidates for explaining the observed dark matter abundance~\cite{Alexander:2016aln,Battaglieri:2017aum}. The null results from many direct detection experiments have put significant constraints on dark matter scenarios where dark matter interacts with the standard model through one of the SM gauge interactions. This motivates the exploration of dark sector models, in which one can still explain the dark matter production through it's interactions with the standard model without being in tension with direct and indirect detection searches.

In dark sector models the SM interacts with dark matter through a weakly coupled SM neutral mediator. The most natural possibilities for such mediators are directly related to the lowest dimensional gauge singlet operators made of SM fields. This singles out three important scenarios: the Higgs, vector, and neutrino portals
\begin{align}
       \delta \mathcal{L}\propto H^{\dagger} H S\,,&& \delta \mathcal{L} \propto J_{\mu}^{\text{SM}} V^{\mu} \,, && \delta \mathcal{L}\propto \bar{L} H N\,,
    \end{align} 
where $S$, $V$, and $N$ are the new scalar, vector, and fermionic fields respectively with $J_\mu^\text{SM}$ representing a current made of SM fields.

The sensitivity of future lepton colliders to these models, and to Dark Matter in general, is already partly explored in the literature \cite{Buttazzo:2018qqp,Ruhdorfer:2019utl,Chacko:2013lna,Ruhdorfer:2024dgz,Ruhdorfer:2023uea,Forslund:2023reu,Li:2024joa,Liu:2017zdh,Han:2020uak,Bottaro:2021snn,Asadi:2023csb,Cesarotti:2024rbh,Capdevilla:2021fmj,Capdevilla:2024bwt,Franceschini:2022sxc,Banerjee:2015gca,Chakraborty:2022pcc,Li:2023tbx,Barik:2024kwv}. 
The sensitivity to Higgs portals has been extensively addressed either by the production of new states \cite{Buttazzo:2018qqp,Ruhdorfer:2019utl,Chacko:2013lna,Ruhdorfer:2024dgz}, or through the branching ratio of the Higgs to invisible \cite{Ruhdorfer:2023uea,Forslund:2023reu,Li:2024joa}. 
Preliminary studies for dark vectors, framed in terms of searches for heavy $Z'$ bosons, can be found in \cite{Karliner:2015tga,Dasgupta:2023zrh,Barik:2024kwv}.\footnote{See also \cite{Liu:2017zdh} for the sensitivity of future Z factories to models featuring vector portal-like interaction to the SM.}
Finally, the sensitivity to neutrino portals has been explored in heavy neutral lepton searches \cite{Banerjee:2015gca,Chakraborty:2022pcc,Li:2023tbx}. In this paper the picture is completed for vector portals by a comprehensive sensitivity study of future lepton colliders to light vector portals, including both visible and invisible decays. 

Precision machines such as lepton colliders offer a unique opportunity to search for weakly coupled states. Furthermore, the clean leptonic environment, with a fixed CoM energy, allows one to precisely measure missing energy, $\slashed{E}$. 
This can be used to search for weakly coupled states, even when they decay invisibly. In this work we will focus on the FCC-ee for our predictions, but most of the results can be easily translated to other $e^+e^-$ colliders running at the same energy.
The large luminosity expected for the FCC-ee will translate to a much greater sensitivity to weakly coupled dark sectors than its predecessor LEP~\cite{Fox:2011fx}. The \(Z\)-pole section of the FCC-ee is aiming to produce $10^{12}$ $Z$-bosons, enabling sensitivity to couplings of order $10^{-3}$.

The advantages of a $\mu$-collider in this area are less apparent, since it targets only percent-level precision. 
However, the large center-of-mass energy ($\sqrt{s}$), combined with a luminosity that scales with energy \cite{Accettura:2023ked}
\begin{equation}
    \mathcal{L} = 10 \textrm{ ab}^{-1}\left(\frac{\sqrt{s}}{10 \textrm{ Tev}}\right)^2,
\end{equation}
enables percent-level precision across a wide range of energies. 
Furthermore, the potential to cover the forward collider region with $\mu$ detectors (see \cite{InternationalMuonCollider:2024jyv}) is particularly promising in this context. A forward detector would allow the exploitation of kinematically favored configurations, such as the enhanced radiation of soft and/or collinear light mediators from $\mu$'s. Finally, the $\mu$-collider has clear advantages in searching for particles that couple preferentially to muons over electrons.
We investigate how these advantages give rise to rich phenomenology and assess the sensitivity of the $\mu$-collider across a broad spectrum of dark vector masses.

The paper is structured as follows. In Sec.~\ref{Sec:Models} we briefly introduce the models under investigation and provide an overview of the various collider and discovery channels analyzed in the rest of the paper. Secs.~\ref{Sec:Invisible} and \ref{Sec:Visible} contain the technical details of our analysis for the invisible and visible channels respectively. Finally, in Sec.~\ref{Sec:Conclusion} we present our results and conclusions. Impatient readers may skip Secs.~\ref{Sec:Invisible} and \ref{Sec:Visible}, and go directly to Sec.~\ref{Sec:Conclusion} for our main findings.

\begin{table}[]
    \centering
    \begin{tabular}{|c||c|c|}
    \hline
        Collider & CoM Energy & Tot. Lum. 
        \\ \hline 
    $\mu$C & 3 TeV& 0.9$\,\text{ab}^{-1}$ \\
    & 10 TeV  &10$\,\text{ab}^{-1}$\\ \hline 
    FCC-ee & 91 GeV (Z-pole) & $6 \times 10^{12}$ Z\\
    & 240 GeV (ZH)& 5 $\,\text{ab}^{-1}$\\ \hline
    \end{tabular} \hspace{0.5cm}
    \begin{tabular}{|c||c|c|}\hline
        Collider & $p_T^l $[GeV] & $|\eta_l|$ \\ \hline 
    $\mu$C (central) & $>10$ & $<2.5 $ \\ \hline
    $\mu$C (forward) & $>10$ & $>2.5,\,<7 $ \\ \hline 
   FCC-ee & $>10^*$ & $<2.5 $ \\ \hline
    \end{tabular}
    \caption{Left: Colliders, energies and integrated luminosity considered for the various sensitivity projections.  Right: Baseline selection criteria for the setups considered in the paper. The forward selection criteria applies only to $\mu$ and includes a further 500 GeV cut on the energy of the forward $\mu$. For some of the FCC-ee analyses we lower the $p_T$ cut on the lepton to $1$ GeV.}
    \label{tab:CollSpec}
\end{table}

\section{Models}\label{Sec:Models}

The focus of this paper is on the so-called vector portals. Vector portals consist of a massive vector, $Z'$, of mass, $M_{Z'}$, coupled to both the SM and some dark sector with the interaction
\begin{align}\label{Eq:GenericGaugePortal}
    \mathcal{L}_{\rm int}= g_{Z'} J_{\mu}^{\text{SM}} Z'^{\mu}+ g_{D} J_{\mu}^{\text{Dark}} Z'^{\mu} \,,
\end{align}
where $J_{\mu}^{\text{SM}}$ is a SM $U(1)$ {global} current. For simplicity, the case where $J_{\mu}^{\text{SM}}$ is a non-anomalous current is considered. For such a current neither a Wess-Zumino term or further assumptions on the dark sector are needed to cancel the anomaly (see \cite{Dror:2017ehi,DiLuzio:2022ziu} for a detailed discussion on this point).

There are various choices for the new gauged \( U(1) \), but we focus only on the dark photon, where $J_\mu^\text{SM}$ is the E$\&$M current, and \( L_{\mu} - L_{\tau} \) portals (with $L_x$ being the lepton number of species $x$). The analysis can be easily extended to other portals. Further assumptions on the dark sector and on the couplings in eq.~\eqref{Eq:GenericGaugePortal} split the analysis into two regimes:
\begin{itemize}
    \item \textbf{Visible decays}: either $g_{Z'} \gg g_{D}$, or the dark sector states are too heavy for the $Z'$ to decay into the dark sector. In this limit the $Z'$ decays exclusively to the SM.
    \item \textbf{Invisible decays}: this is achieved for $g_{D} \gg g_{Z'}$ and assuming there are light invisible states (i.e. not decaying to visible SM particles) in the dark sector. We thus neglect the branching ratio to the SM and only consider missing energy, $\slashed{E}$, as the final state for collider searches.
\end{itemize}

While we only focus on these two models we emphasize that those cover all the phenomenology for $Z'$ searches at future lepton colliders. Indeed our sensitivity projections can be readily translated to any gauge boson with $L_{e}$ and/or $L_{\mu}$ charges.

\subsection{Dark photon portal}

The first model is the so-called dark photon or hypercharge portal (see \cite{Graham:2021ggy} for an extensive review). 
A vector of mass, $M_{Z'}$, kinetically mixes with the hypercharge gauge boson. The interactions between the SM and the dark sector are
\begin{align}\label{Eq:DarkPhotonL}
\mathcal{L}_{\text{int}} =  \frac{\epsilon}{2 c_W} Z_{\mu \nu}' B^{\mu \nu} + g_D Z'_{\mu} J^{\mu}_{\text{Dark}}\,,
\end{align}
where $Z_{\mu \nu}'$, $B_{\mu \nu}$ are the fields strengths for the dark photon and the SM $B$ field respectively, $\epsilon \ll 1$ is the kinetic mixing parameter, and $c_W = \cos \theta_W$ with $\theta_W$ the Weinberg mixing angle.

The interactions in eq.~\eqref{Eq:DarkPhotonL} can be brought into the form of eq.~\eqref{Eq:GenericGaugePortal} by a suitable field redefinition diagonalising both the kinetic and mass terms.
For details of the diagonalisation procedure consult \cite{Curtin:2014cca,Harigaya:2023uhg}. 
For this work the interactions between the dark photon and the fermionic currents are considered in the limit $\epsilon \ll |1-r|$\footnote{This approximation breaks down in a tiny region of a few GeV around $M_{Z'}=M_Z$. That window is not to be considered reliable in our sensitivity projections.}
\begin{align}\label{Eq:DPInteractions}
  \mathcal{L}_{\rm int} = \frac{e \epsilon}{1-r} Z'_\mu\left(J^{\mu}_{em} - \frac{r}{c_W^2} J^{\mu}_{Y}\right)+g_D Z'_{\mu} J^{\mu}_{\text{Dark}}\,,
\end{align}
where $r= M_{Z'}^2 / m_Z^2 $, $e$ is the electric charge of the election, and $J_{\text{em}}$, $J_{Y}$ are the SM electromagnetic, hypercharge currents respectively. 
Note that if $M_{Z'} \ll M_Z$, this leads to $g_{Z'} \approx e \epsilon$, while for $M_{Z'} \gg M_Z$ this leads to $g_{Z'} \approx  e \epsilon / c_W^2$. Thus, for $\epsilon \ll 1$ in this model it is generically expected that $g_D > g_{Z'}$. Whether the dark photon decays visibly or invisibly is therefore determined by whether there are light dark sector states.

Dark photons are targets for a plethora of searches. In the mass range of interest ($m_{Z'} \gtrsim 10$ GeV) collider searches are particularly relevant, either through resonance searches (visible scenario) or via missing energy searches (invisible scenario). 
FCC-ee and the $\mu$-collider offer a rich phenomenology through several production channels, as summarized in Tab.~\ref{tab:Zp_channel}. 

For invisibly decaying dark photons the main constraints arise from mono-$\gamma$ searches at LEP \cite{Fox:2011fx}, which constrain $\epsilon \gtrsim 10^{-1} - 10^{-2}$ for $M_{Z'} \gtrsim 10$ GeV, and EWPT \cite{Curtin:2014cca}. Mono-$\gamma$ \cite{ATLAS:2020uiq} and mono-jet \cite{CMS:2021far} searches at the LHC are less sensitive. Constraints from BaBar and the expected sensitivity from Belle-II~\cite{Graham:2021ggy} are significantly stronger but only apply for sufficiently light dark photons, $M_{Z'} \lesssim 10$ GeV. 

Future lepton colliders can open up new regions of parameter space. The FCC-ee, being basically a more precise version of its predecessor, can improve bounds by least one order of magnitude in $\epsilon$ based on its expected luminosity (see Tab.~\ref{tab:CollSpec}). In contrast, the $\mu$-collider aims for a precision similar to LEP-II, but covers a much wider range of masses. Furthermore, the larger CoM energy offers a richer phenomenology than a lepton collider operating around the EW scale.\footnote{The $\mu$-collider is also extremely powerful in testing $Z'$ through indirect searches for higher-dimensional operators, that pushes the sensitivity up to $m_{Z'} \sim 100$ TeV \cite{Chen:2022msz}. Interestingly, a $\mu$ beam dump experiment at the $\mu$-collider can further complement these searches and explore new regions of parameter space in the MeV-GeV mass range \cite{Cesarotti:2022ttv}.}

In the visible scenario LHCb \cite{LHCb:2019vmc} and CMS \cite{CMS:2024zqs,CMS:2021ctt} searches for di-muon resonances are particularly strong, covering a mass range from 1 GeV to a few TeV. Electroweak precision tests (EWPT) \cite{Curtin:2014cca} are relevant for masses around the $Z$-boson mass, in the region where large $Z$-backgrounds make resonance searches ineffective. All these searches essentially exclude any dark photon with $\varepsilon \gtrsim 10^{-3}$ in the relevant mass range. 
As we will show, only FCC-ee has the precision to compete with current bounds, either through searches for exotic $Z$ decays (for $M_{Z'} \lesssim M_Z$) or through mono-$\gamma$ searches. The $\mu$-collider, although targeting percent-level precision, is competitive for masses above approximately 2 TeV, where the LHC bounds become weak due to low effective luminosity.

\begin{table}[t]
    \centering
    \begin{tabular}{|c|c|c|c|}
        \hline 
         \multirow{2}{*}{Collider} & \multirow{2}{*}{Production} & \multicolumn{2}{c|}{Background} \\ 
         && Invisible decay & Visible decay \\ \hline 
         \multirow{3}{*}{$\mu$C} & $\mu^+ \mu^- \rightarrow Z' \gamma$ & $\mu^+ \mu^- \rightarrow \nu_l \bar{\nu_l} \gamma$ & $\mu^+ \mu^- \rightarrow \mu^+ \mu^- \gamma$ \\ 
         & \multirow{2}{*}{$\mu^+ \mu^- \rightarrow Z' \mu^+ \mu^-$} & $\mu^+ \mu^- \rightarrow \gamma (\slashed{E}) \mu^+ \mu^-$ & \multirow{2}{*}{$\mu^+ \mu^- \rightarrow 4 \mu$} \\ 
         &  & $\mu^+ \mu^- \rightarrow \nu_l \bar{\nu_l} \mu^+ \mu^-$ & \\ \hline 
         \multirow{2}{*}{FCC-ee (Z pole)} & $Z \rightarrow Z' \mu^+ \mu^-$ & $Z \rightarrow \tau \tau (\tau \rightarrow \mu \nu_{\mu} \nu_{\tau})$ & $e^+ e^- \rightarrow 4 \mu$ \\ 
          &$ \dagger$ \, $e^+e^- \to Z' \gamma$ & $e^+e^- \to \nu_l \bar{\nu_l} \gamma$ & $e^+e^- \to \mu^+\mu^- \gamma$\\ \hline
         FCC-ee (Zh) & $\dagger$ \, $e^+ e^- \rightarrow Z' \gamma$ & $e^+ e^- \rightarrow \nu_{l} \bar{\nu}_{l} \gamma$ & $e^+ e^- \rightarrow \mu^+ \mu^- \gamma$ \\ \hline
    \end{tabular}
    \caption{The $Z'$ (dark photon or $L_{\mu} - L_{\tau}$) production channels considered in the text, along with the main backgrounds. Entries marked by a ``$\dagger$'' apply only to the dark photon portal. The second production channel for the $\mu$-collider potentially benefits from forward $\mu$ detectors, as explained in the main text.}
    \label{tab:Zp_channel}
\end{table}

\subsection{$L_{\mu}-L_{\tau}$ portal}

The second model is the so-called $L_{\mu}-L_{\tau}$ portal. In this case the SM current in eq.~\eqref{Eq:GenericGaugePortal} is associated with the $L_{\mu}-L_{\tau}$ charge, namely
\begin{align}\label{Eq:LmuLtauCurrent}
 J^{\text{SM}}_{\rho} =  \bar{\mu}\gamma^{\rho} \mu - \bar{\tau} \gamma^{\rho} \tau + \bar{\nu}_{\mu} \gamma^{\rho} \nu_{\mu}-\bar{\nu}_{\tau} \gamma^{\rho} \nu_{\tau}\,.
\end{align}
As before the $Z'$ is assumed to have mass $M_{Z'}$. We are agnostic about the origin of its mass, but note that it can lead to additional constraints~\cite{Ekhterachian:2021rkx}.
 
This simple extension of the SM is tested by a variety of probes: dark matter direct detection \cite{Harnik:2012ni}, exotic $W$ decays \cite{Agashe:2023itp,Fei:2024qtu}, K- factories \cite{Krnjaic:2019rsv}, and neutrino beam dump experiments \cite{Altmannshofer:2014pba}. Beam dumps constrain $g_{Z'}\lesssim 10^{-2} - 10^{-1}$ for both invisibly and visibly decaying $Z'$ in the mass range of interest, $M_{Z'}\sim 1-100$ GeV.  Furthermore, the latter is tested by the LHC up to couplings as small as $10^{-2}$ for masses $\lesssim\, 50$ GeV through the $4\mu$ channel \cite{ATLAS:2023vxg,CMS:2018yxg}. 

A $\mu$-collider clearly represents an ideal probe for this specific model, where new physics couples directly to muons. 
The new $Z'$ can be produced in two different channels in a $\mu$-collider (outlined in Tab.~\ref{tab:Zp_channel}) for a wide range of masses, from a few GeV to the CoM energy (see also \cite{Dasgupta:2023zrh,Barik:2024kwv}). 
The FCC-ee, by precisely determining $Z$-decays, is promising for masses below the $Z$-boson mass.
The associated production with a photon at FCC-ee, which is the most sensitive channel in a large range of masses for the dark photon, does not apply to the $L_{\mu}-L_{\tau}$ portal.

\section{Decays to the hidden sector}\label{Sec:Invisible}

We start our detailed analysis with the invisibly decaying $Z'$. The various discovery channels for both the $\mu$C and the FCC-ee are listed in Table~\ref{tab:Zp_channel}.

\subsection{Muon collider}

\begin{figure}[t]
  \includegraphics[width=\textwidth]{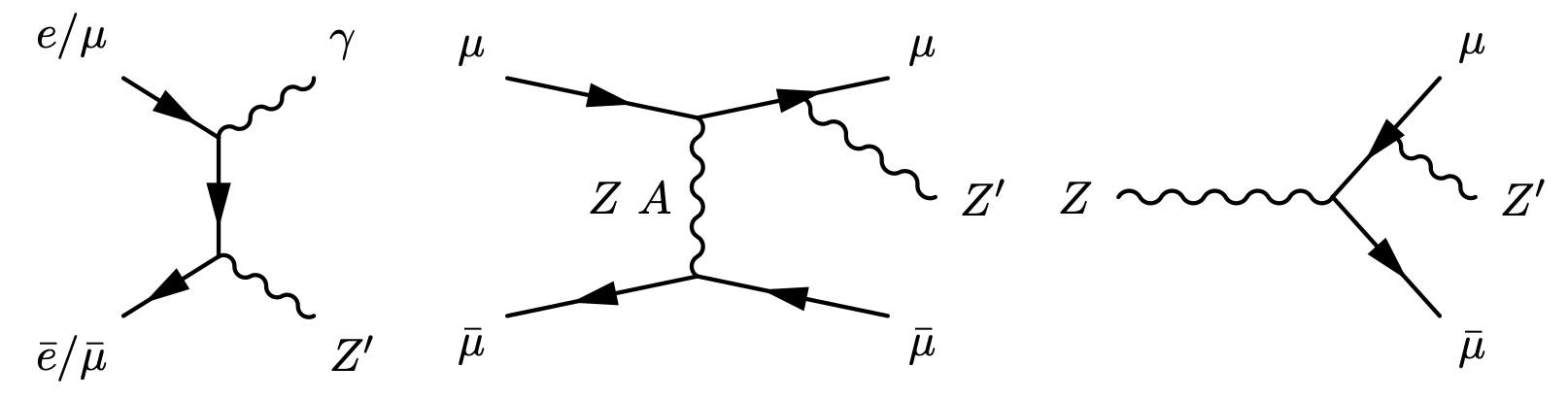}
  \caption{Representative diagrams for vector portals at future lepton colliders. Left: associated production with a photon. Middle: $Z'$ bremsstrahlung. Right: Exotic $Z$ decays. }
    \label{fig:MuoCZpProd}
\end{figure}

The two channels relevant at the $\mu$C (see Tab.~\ref{tab:Zp_channel}) are the associated production and the $Z'$-bremsstrahlung depicted by the left and middle Feynman diagrams in Fig.~\ref{fig:MuoCZpProd} respectively. We will consider two CoM energies, 3 and 10 TeV, with luminosities given in Table~\ref{tab:CollSpec}. 
As discussed in the introduction, the $Z'$-bremsstrahlung from $t$-channel $\mu^+\mu^-$ scattering is enhanced in the forward region and can potentially benefit from a forward $\mu$-detector. 
Thus, only for this channel, we will study the impact of the forward detector on the sensitivity projections. 

Monte-Carlo simulations are used for all predictions. 
Events for all signal and background processes are generated at leading order in \texttt{MadGraph 5} \cite{Alwall_2011}, then showered through \texttt{Pythia 8} \cite{Sj_strand_2015}.\footnote{Our simulations do not include Initial State Radiation showering since no public implementation of Parton Distribution Functions for muons is available in \texttt{MadGraph 5}. ISR effects can become important for low $p_T$ processes (see for instance \cite{Ruhdorfer:2023uea}). However, such events are excluded by our selection criteria. A more refined analysis, including a detailed treatment of the radiative corrections, is beyond the scope of out work.}
Baseline detector effects are simulated using the muon collider card in \texttt{Delphes 3} \cite{Delphes2014}. Additional detector effects are included in the forward region as detailed below. All the selected events pass the minimal selection criteria in Table~\ref{tab:CollSpec} unless otherwise specified.

\subsubsection*{Associated Production}

First, we consider the associated production of a $Z'$ with a hard photon where the $Z'$ decays invisibly (see left panel of Fig.~\ref{fig:MuoCZpProd}).
The primary background in this case is $ \gamma\, \nu_l \, \bar{\nu}_l$. Other backgrounds, such as $\gamma \, \gamma$ with one photon undetected, are subdominant. Signal events are characterized by the energy of the photon, $E_\gamma$, which can be used to reconstruct the mass, $M_{Z'}$, using the following leading order relation
\begin{equation}\label{eq:monop_mzprecon}
    E_\gamma = \frac{s - M_{Z'}^2}{2\sqrt{s}},
\end{equation}
with $\sqrt{s}$ the CoM energy. 

\begin{figure}[t]
    \centering
\includegraphics[width=0.9\linewidth]{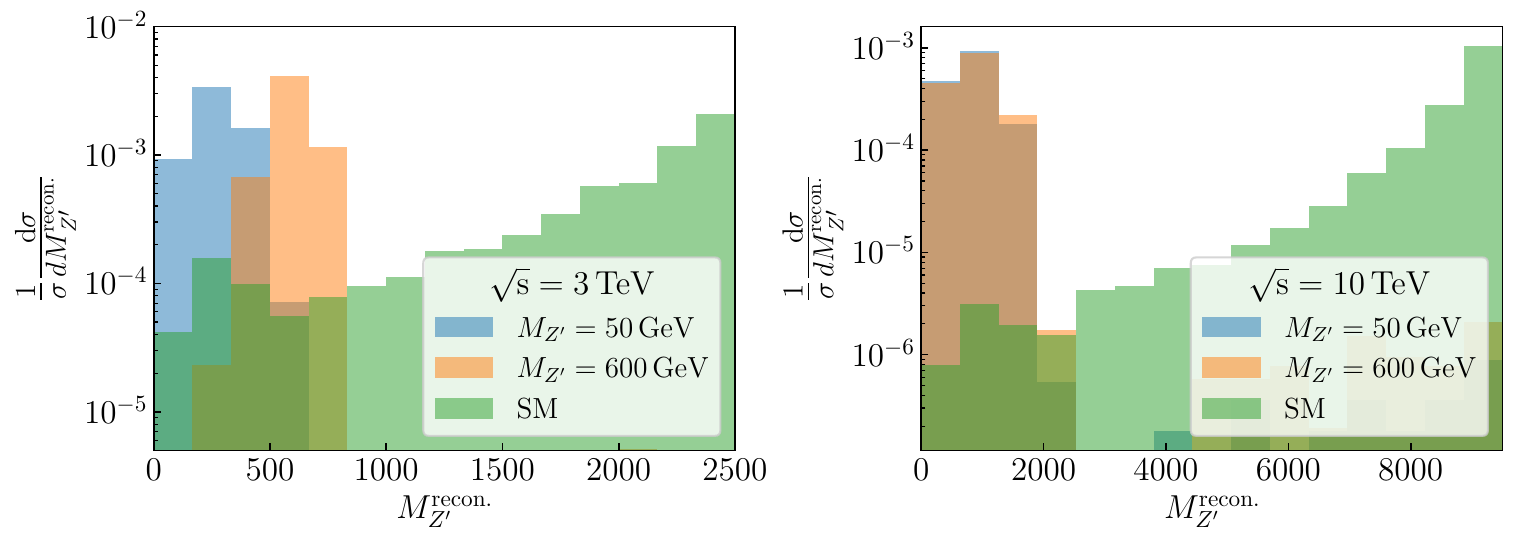}
    \caption{Example normalized differential cross-section for the process $\mu^+ \mu^-  \rightarrow Z' (\slashed E) \, \gamma $ as a function $M_{Z'}^{\rm{recon.}}$. The left, right panel corresponds to a $3,10$ TeV collider respectively, and $\epsilon =0.03$ is chosen.}
    \label{fig:monop_insvisible}
\end{figure}

The reconstructed invariant mass in Eq.~\eqref{eq:monop_mzprecon} can be significantly smeared by finite detector resolution when $\sqrt{s}\gg M_{Z'}$. 
A small error $\Delta E_\gamma$ in the measurement of the photon energy translates to a large error $\Delta M_{Z'}^{\rm{recon.}}$ in the reconstructed mass $M_{Z'}^{\rm{recon.}}$   
\begin{equation}\label{eq:monop_smear}
    \frac{\Delta M_{Z'}^{\rm{recon.}}}{M_{Z'}^{\rm{recon.}}} \sim  \frac{s}{2 M_{Z'}^2}\frac{\Delta E_\gamma}{E_\gamma}\,.
\end{equation}
We show example normalised reconstructed mass distributions for signal and background in Fig.~\ref{fig:monop_insvisible} for two benchmark points in the dark photon portal with the two considered collider energies. 
To separate signal from background we impose cut-offs of:
\begin{align*}
    &|M_{Z'}^{\rm{recon.}} - M_{Z'}|<300\,\text{GeV},\hspace{1cm} \sqrt{s}=\text{3 TeV},\\
    &|M_{Z'}^{\rm{recon.}} - M_{Z'}|<1\,\text{TeV} \,,\hspace{1.3cm}\sqrt{s}=\text{10 TeV}.
\end{align*}
This selection removes most of the background for $Z'$ masses in the range of interest. 

The $95\%$ CL sensitivity projections are shown in Fig.~\ref{fig:Inv_Bound} (dotted lines). The left and right panel show respectively the parameter space in the dark photon and $L_{\mu}-L_{\tau}$ portals. We can see that in both scenarios the sensitivity would be significantly improved over present constraints. Furthermore, notice that no notable improvement is expected by increasing the CoM energy, since the expected $\mu$C luminosity ($\mathcal{L}\sim s$) keeps the number of events in two-to-two processes constant.

\subsubsection*{$Z'$-bremsstrahlung}
The second channel we study is the $Z'$-bremsstrahlung from a muon in $\mu^+ \,\mu^-$ scattering (see middle panel of Fig.~\ref{fig:MuoCZpProd}). For this channel we also study the effect of 
installing a forward $\mu$ detector that covers the pseudo-rapidity region $2.5<|\eta|<7$ since the cross-section is enhanced at low momentum transfer.
We assume this detector to have a nominal $10\%$ energy resolution and an angular resolution of $5$ mrad.\footnote{The $10\%$ energy resolution is already implemented in the \texttt{Delphes} muon collider card. On top of that we add a gaussian random variable centered at zero and standard deviation 5 mrad to the azimuthal and polar angle to smear $\eta$ and $\phi$. Additional results with 0 and 50 mrad can be found in App.~\ref{App:AddResults}.}
A $500$ GeV cut is imposed on the energy of the forward muons otherwise they do not make it to the forward detector. 
These parameters are in line with previous phenomenological studies \cite{Ruhdorfer:2023uea}.

\begin{figure}[t]
    \centering
    \includegraphics[width=1\linewidth]{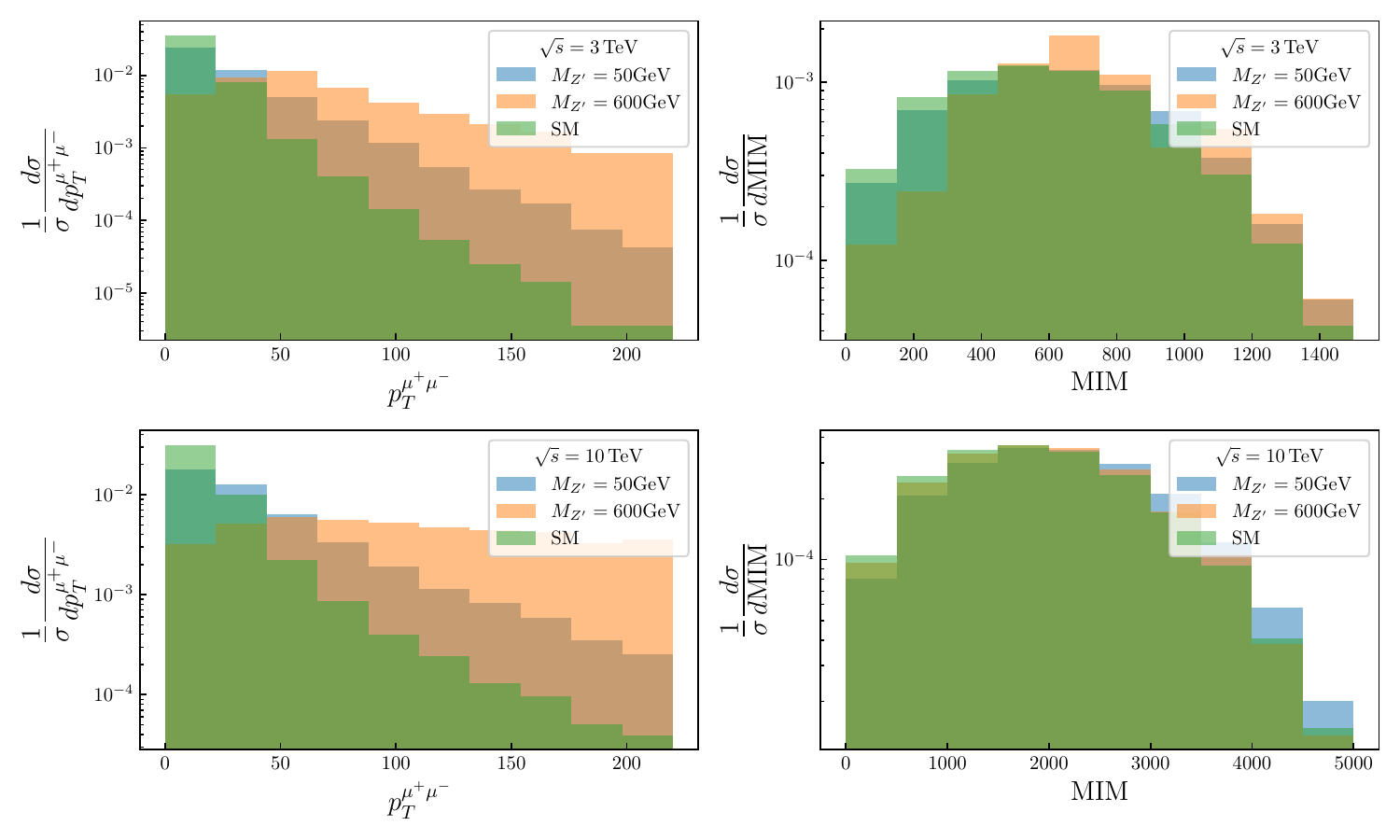}
    \caption{Example normalized differential cross-section for the process $\mu^+ \mu^-  \rightarrow Z' (\slashed E) \, \mu^+ \, \mu^- $ as a function as function of Missing Invariant Mass (MIM) and $p_T^{\mu^+ \mu^-}$  for $3$ TeV (top) and $10$ TeV (bottom) collider, $\epsilon =0.03$. The figure just shows the forward $\mu$ detectors ($2.5<\eta<7$). 
    }
    \label{fig:forward_muon_spec}
\end{figure}

In this scenario the final state we consider is two muons of opposite charge and missing energy. The dominant backgrounds are thus $\mu^+ \mu^- \nu_l \bar{\nu_l}$, and $\mu^+ \mu^- \gamma$ with the final state photon escaping detection, usually due to being forward and/or soft. In principle, we can reconstruct the missing invariant mass (MIM)
\begin{align}\label{Eq:MIMRareZdecays}
    \text{MIM}_{Z'} = \sqrt{\left(p_i- p_{\mu^+} - p_{\mu^-} \right)^2}\,,
\end{align}
where $p_{\mu^-}$, $p_{\mu^+}$ are the 4-momenta of the final muons and $p_i$ is the total ingoing 4-momenta. However, reconstruction of MIM is impractical for light masses compared to the collider energy $M_{Z'} \ll \sqrt{s}$ because of the finite energy resolution and beam energy spread \cite{Ruhdorfer:2023uea,Li:2023tbx}. This is shown in the right panels of Fig.~\ref{fig:forward_muon_spec} for two benchmark values. Notice that in the Figure we just show these effects in forward $\mu$-detectors.

On the other hand, the transverse momentum of the $\mu^+ \mu^-$ pair, $p_T^{\mu \mu}$, for the signal is harder than that of the background (see Fig.~\ref{fig:forward_muon_spec}) and can thus be exploited to enhance the sensitivity as pointed out in similar phenomenology studies \cite{Ruhdorfer:2023uea,Li:2024joa}.
Hence we impose a minimum $p_T^{\mu\mu}$ cut, optimized over the $M_{Z'}$ range, to enhance the sensitivity.\footnote{The minimal cut we impose is $p_T^{\mu\mu}>50$ GeV, which also mitigates large radiative effects from collinear photon emissions, reducing the theoretical uncertainty of our fixed-order calculation. See \cite{Ruhdorfer:2024dgz, Ruhdorfer:2023uea, Li:2024joa} for further details.} Finally, we point out that different selection criteria can further enhance the sensitivity to heavier masses, however, at higher masses, associated production provides more stringent constraints than the radiated $Z'$ production channel discussed here.

The $95\%$ CL sensitivity projections are given in Fig.~\ref{fig:Inv_Bound}, including the results with (solid lines) and without (dashed lines) a forward detector. The forward detector leads to a mild improvement for sufficiently light dark photon masses for which the cross-section is enhanced when the final state muons are in the forward region. Furthermore, notice that there is a mild improvement with increased collider energy, in contrast to the associated production channel. In fact, while the number of events in two-to-two processes are constant at the $\mu$C, radiation is naturally enhanced at higher energies.

\subsection{FCC-ee}

We now turn to the projected sensitivity for the FCC-ee. The collaboration is currently considering various stages of the project, each with different energies and luminosities \cite{FCC:2018evy}. For our study we focus on the two key pillars of the FCC-ee: the \(Z\) factory and the Higgs factory (see Tab.~\ref{tab:CollSpec}).

The Z factory presents a promising opportunity to test invisibly decaying gauge bosons that are considerably lighter than the SM Z boson by carefully analyzing exotic decay channels (see right panel of Fig.~\ref{fig:MuoCZpProd}). 
Additionally, both the Higgs and $Z$ factory stages have the potential to probe a $Z'$ boson via associated production with a photon (see left panel of Fig.~\ref{fig:MuoCZpProd}), provided that it couples directly to electrons. This is the case for the dark photon, but not for the \(L_{\mu} - L_{\tau}\) model. Furthermore, this process is directly analogous to the process studied at LEP II, which continues to set the most stringent constraints on the dark photon portal across a substantial region of parameter space \cite{Fox:2011fx}. 

All Monte Carlo events for these analyses are generated at leading order using \texttt{MadGraph 5} \cite{Alwall_2011}, with events required to meet only the minimal selection criteria outlined in Tab.~\ref{tab:CollSpec}, unless otherwise indicated. Additional detector effects, such as finite energy resolution and beam energy spread, are incorporated where relevant.

\subsubsection*{Associated production}

We begin with the associated production of a $Z'$ which mostly decays to the dark sector. The final state is $\gamma \slashed{E}$ with primary background $\gamma \nu_l \bar{\nu}_l$. To include minimal detector effects in our analysis, we start from tree-level events and smear the energy of the photon in the final state by $4\%$.\footnote{This value is in line with the expected performances for the CLD calorimeter \cite{FCC:2018evy}.} Then, we use Eq.~\eqref{eq:monop_mzprecon} to reconstruct the $Z'$ mass ($M_{Z'}^{\rm{recon.}}$) for the signal and background events. The reconstructed mass is significantly smeared by detector effects (see Eq.~\eqref{eq:monop_smear}), especially for light $Z'$ masses. Thus, on top of the baseline selection of Tab.~\ref{tab:CollSpec}, we apply the conservative selection criteria:
\begin{align*}
&|M_{Z'}^{\rm{recon.}}-M_{Z'}|< \left(\frac{40 \text{ GeV}}{M_{Z'} }\right )^2 8 \text{ GeV},\hspace{1cm}\text{Z-Pole},\\ &|M_{Z'}^{\rm{recon.}}-M_{Z'}|< \left(\frac{50 \text{ GeV}}{M_{Z'}}\right )^2 20\text{ GeV},\hspace{0.8cm}\text{ZH}. 
\end{align*}
We have chosen these values based on Eq.~\eqref{eq:monop_smear} and checked its consistency numerically. In the left panel of Fig.~\ref{fig:Inv_Bound} we present the 95\% exclusion limits. The dashed and solid lines correspond to the bounds obtained with and without reconstructing the invariant mass respectively. First we observe two sensitivity enhancements due to resonant $Z'-Z$ mixing and radiative return. Additionally, we see a loss in sensitivity for the solid line at lower $Z'$ masses, which is attributed to reduced control over the invariant mass reconstruction (see Eq.\eqref{eq:monop_smear}).

\subsubsection*{Exotic Z decays}
For $Z'$ lighter than the $Z$ searches for exotic $Z$-decays at FCC-ee become a sensitive probe of the vector portal scenario. In particular we consider the decays
\begin{align}
    e^+ e^- \rightarrow Z \rightarrow \mu^+ \mu^- Z' (\slashed{E})\,,
\end{align} 
with the corresponding background given in Tab.~\ref{tab:Zp_channel}. In order to maximize the sensitivity we lower the $p_T$ cut on muons down to $1\text{GeV}$. We reconstruct $M_{Z'}$ from the Missing Invariant Mass as in Eq.~\eqref{Eq:MIMRareZdecays}. To estimate finite detector effects we include $0.13\%$ Beam energy spread and $0.1$ GeV smearing on each component of the muon momenta \cite{FCC:2018evy}. 

The 95\% CL projections are shown by the red lines in the right panel of Fig.~\ref{fig:Inv_Bound} and the yellow line in the left panel.
The dashed line is obtained by considering the total number of events passing the cut selection. 
The solid red line shows the projection when using the reconstructed MIM of the events. 
Using the spread of MIM in the simulated signal events (solid red line), we identify the range of MIM centered at the simulated mass containing $95\%$ of the signal events. In the analysis using the reconstructed invariant mass, we only select events that fall within this $95\%$ region. 
Notice that in this process we assume the invisible width of the $Z'$ is always smaller than the experimental spread for the MIM. This corresponds to a width of at most $\sim$5\% for the largest considered masses ($M_{Z'}\sim 70$ GeV) and $\sim$10\% for the lightest ($M_{Z'}\sim 10$ GeV). For the dark photon the exotic decay channel is the most sensitive channel in a small range of masses, and only by employing the MIM reconstruction. Thus, in Fig.~\ref{fig:Inv_Bound} we only report the stronger projection obtained by reconstructing the MIM as described above.

\section{Decays to the Standard Model}\label{Sec:Visible}

We now consider the visibly decaying case where all $Z'$s decay into SM particles. 
As in the previous discussion, we detail our analyses in Sec.~\ref{Sec:MuonCollVisible} and Sec.~\ref{Sec:FCCeeVisible} respectively for the $\mu$C and the FCC-ee. The various discovery channels are listed in Table~\ref{tab:Zp_channel}. Discussion of the results can be found in Sec.~\ref{Sec:Conclusion}.

\subsection{Muon collider}\label{Sec:MuonCollVisible}
\subsubsection*{Associated production}

We consider the production of a $Z'$ with a photon (see the left panel of Fig.~\ref{fig:MuoCZpProd}). Our focus is on $Z'$ decaying primarily to the SM, and specifically to $\mu^+ \mu^-$. The primary background for this process is $\mu^+\, \mu^- \, \gamma$. Notice that additional channels, such as $e^+e^-$, could be relevant for dark photon searches. However, for simplicity we concentrate on decays to $\mu-$pairs, though additional channels would modestly improve our projections.\footnote{We numerically checked that backgrounds in the region of interest are comparable for electron and muon final states.}

\begin{figure}[t]
    \centering
\includegraphics[width=0.99\linewidth]{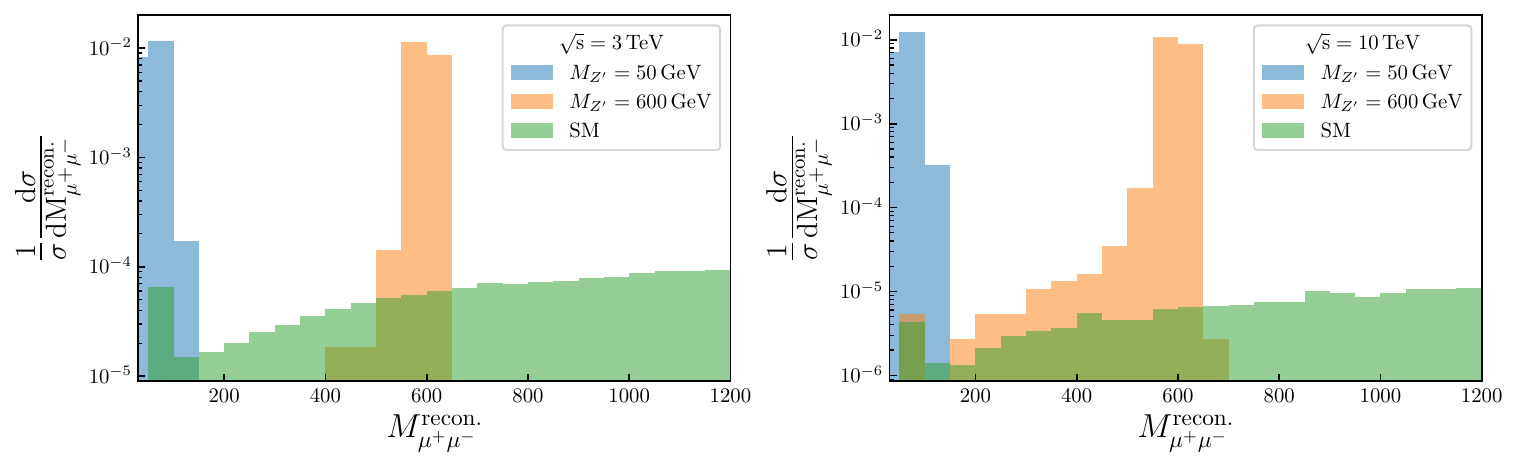}
    \caption{Example normalized differential cross-section for the process $\mu^+ \, \mu^- \rightarrow Z' (\mu^+\, \mu^-)\, \gamma$ as a function of $M_{\mu^+\mu^-}^{\rm{recon.}}$ for $3$ TeV (left) and $10$ TeV (right) collider, $\epsilon =0.03$.}
    \label{fig:monop_visible}
\end{figure}

To minimize the background we select only events where the $\mu^+ \mu^-$ invariant mass, $M_{\mu^+\mu^-}$, is close to the $Z'$ mass $M_{Z'}$. We thus impose a selection cut on the reconstructed di-muon invariant mass 
\begin{equation}|M_{\mu^+\mu^-}-M_{Z'}|<50 \, \rm{GeV}\,.
\end{equation}

This selection is designed to approximately optimize sensitivity, and account for smearing effects due to finite detector resolution, which is a much larger effect than the finite width of the $Z'$. Example normalized distributions of $M_{\mu^+\mu^-}$ can be found in Fig.~\ref{fig:monop_visible} for both $3$, and $10$ TeV $\mu$-colliders.

The deduced $95\%$ CL sensitivity lines shown by the dotted lines in Fig.~\ref{fig:Vis_Bound}.
The blue, cyan lines correspond to a $3,10$ TeV collider respectively, with corresponding luminosities given in Tab.~\ref{tab:CollSpec}.
Notice that there is little change in going from $3$ to $10$ TeV. This behaviour is expected due to the $\mu$-collider's growth in luminosity with increased energy leading to a comparable number of events. 
For this analysis we didn't consider the forward $\mu$ detectors. 
Furthermore, we can see that for masses around the $Z$-boson mass, the mixing in eq.~\eqref{Eq:DPInteractions} enhances the coupling of the Dark photon to the SM current, causing the visible dip in the bounds in Fig. \ref{fig:Vis_Bound}.

\subsubsection*{$Z'$-bremsstrahlung}
Next we study the $Z'$-bremsstrahlung from a muon in $\mu^+ \,\mu^-$ scattering (see middle panel of Fig.~\ref{fig:MuoCZpProd}). For this process we also study the effect of installing a forward $\mu$ detector, as described above. For simplicity we only consider the decay of the $Z'$ to a pair of muons. We estimated that the inclusion of other channels, such as $\mu^+ \mu^- Z' \rightarrow \mu^+ \mu^- e^+ e^-$ for the case of the dark photon, will only marginally improve the overall reach. 

\begin{figure}[t]
    \centering
    \includegraphics[width=1\linewidth]{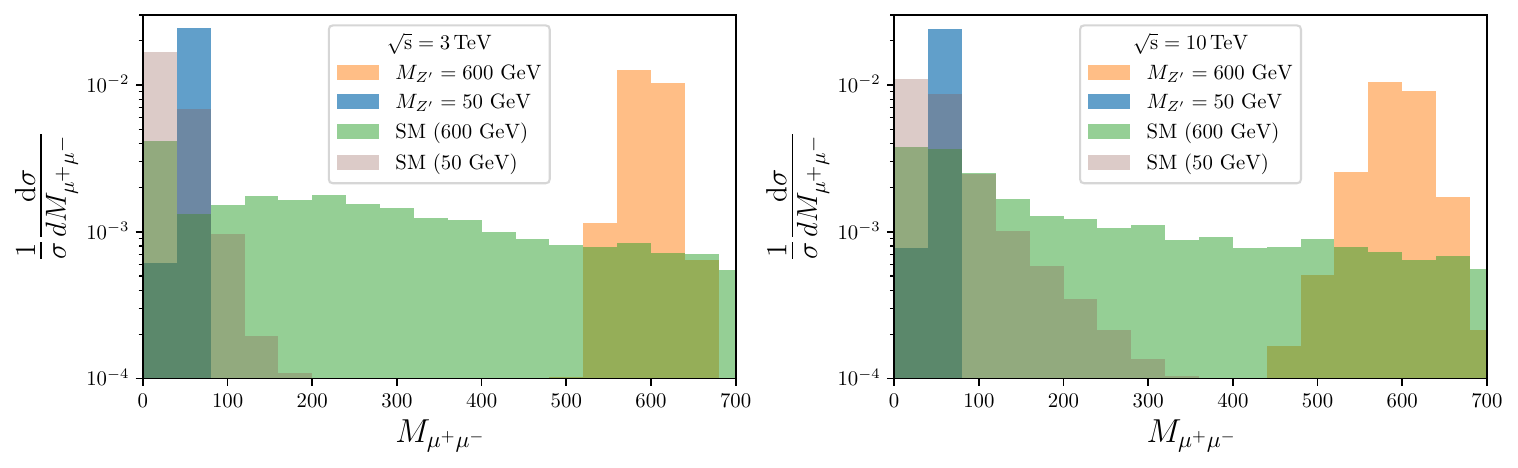}
    \caption{Example normalized differential cross-sections for the process $\mu^+ \, \mu^- \rightarrow Z' (\mu^+\, \mu^-)\, \mu^+\, \mu^-$. The $x$-axis, $M_{\mu^+\mu^-}$, is the invariant mass of the muon pair closest to $M_{Z'}$. We chose  $\epsilon =0.03$, and CoM energies $3$ TeV (left) and $10$ TeV (right). In this figure we just consider forward $\mu$-detectors.}
    \label{fig:radzvisible}
\end{figure}

To minimize backgrounds we only select events with $4\mu$'s passing the baseline selection criteria of Tab.~\ref{tab:CollSpec}. We then construct the invariant mass $M_{\mu^+\mu^-}$ for all 4 possible $\mu^+ \mu^-$ pairs, and we select events where at least one pair satisfies $|M_{\mu^+\mu^-}-M_{Z'}|< 50\, \text{GeV}$. This selection keeps most of the signal events accounting for the finite $Z'$ width and detector smearing while rejecting the background as much as possible (see Fig.~\ref{fig:radzvisible}).

From the selected events we extract the sensitivity projections in Fig.~\ref{fig:Vis_Bound}. The solid lines are the $95\%$ CL exclusion limits including a forward detector, whereas the dashed lines cover only the central region, $|\eta|<2.5$. The inclusion of a forward detector was not found to give a noticeable improvement. Additional projections with variations of the angular detector resolution for the forward detector can be found in App.~\ref{App:AddResults}.

\subsection{FCC-ee}\label{Sec:FCCeeVisible}
\subsubsection*{Associated Production}
In this section we consider production of $Z'$ in association with a photon. We only study the decay of $Z'$ to muons, i.e.,
\begin{equation}
    e^+ e^- \rightarrow Z' \gamma \rightarrow \mu^+ \mu^- \gamma.
\end{equation}
The main backgrounds for this process are $Z$ decay to $\mu^+ \mu^- \gamma$ for the $Z$-pole run and $Z\, \text{or}\, \gamma^* (\mu^+ \mu^-)+ \gamma$ for the $ZH$ run.  This process is only viable for the dark photon since the $L_\mu - L_\tau$ has no direct couplings to electrons.

For our sensitivity projections we impose the selection cuts as described in Tab.~\ref{tab:CollSpec}. Furthermore, we impose a $p_T >$ 1 GeV and $p_T >$ 10 GeV cut on the photon for the $Z-$pole and $ZH$ runs respectively. We further assume to be able to reconstruct the invariant mass of the $Z'$ to within $1\%$ or $0.1\%$, and only select background events in these windows. These values are in line with the expected experimental sensitivity for the CLD detector at FCC-ee \cite{FCC:2018evy}. Note that the width of the $Z'$ is well contained in those values. The results of our analysis are shown in Fig.~\ref{fig:Vis_Bound}.

\subsubsection*{Exotic $Z$ decays}
Finally, the FCC-ee is sensitive to \( Z' \) lighter than the \( Z \) boson through tests involving the exotic decays of the latter. In particular we consider $Z$-boson decays into $4\mu$, namely:
\begin{align}
    e^+ e^- \rightarrow Z \rightarrow \mu^+ \mu^- Z' \rightarrow \,  \mu^+ \mu^- \,\mu^+ \mu^- \,,
\end{align}
with background mostly from direct $Z$ decays to the same final state.

Most of the signal events are associated with the production of a $Z'$ which is relatively soft. Therefore, for this analysis, it is crucial to consider $\mu$ with $p_T$ as low as $1\,\text{GeV}$. As in the previous section, we assume two benchmark values for the experimental resolution on the di-muon invariant mass: 1\% and 0.1\%. For background events, we consider the four possible pairings of the $\mu^+\mu^-$ and select events in which one of the pairings is within $1\%$ (or $0.1\%$) of the selected mass. The sensitivity projections are reported in Fig.~\ref{fig:Vis_Bound} for both the dark-photon and $L_{\mu}-L_{\tau}$. Notice that some additional sensitivity to the former can be gained by considering additional final states.

\section{Results \& Conclusion}\label{Sec:Conclusion}

In this paper we assessed the expected sensitivity of future lepton colliders, namely the FCC-ee and the $\mu$-collider (see Tab.~\ref{tab:CollSpec}), to weakly coupled gauge portals. We specifically focused on the dark photon and $L_{\mu} - L_{\tau}$ portals with couplings given in Eq.~\eqref{Eq:DarkPhotonL} and Eq.~\eqref{Eq:LmuLtauCurrent} respectively. For both models we considered two cases: the $Z'$ may decay primarily into a hidden sector (invisibly decaying case), or the $Z'$ may decay primarily into the Standard Model (visibly decaying case). In addition to this, we emphasize that for all practical purposes our results depend only on the $L_{e}$ and $L_{\mu}$ charges of the $Z'$s, and can thus be easily adapted to any vector portal coupling to either electrons or muons.

We explored the production channels listed in Tab.~\ref{tab:Zp_channel} for both the visible and invisible cases. The technical details of our analysis are provided in Sec.~\ref{Sec:Invisible} and Sec.~\ref{Sec:Visible}. We summarize our main findings below.

\subsection{Decays to hidden sector}

\begin{figure}
\begin{minipage}{0.45\linewidth}
    \centering
    \includegraphics[width=1\linewidth]{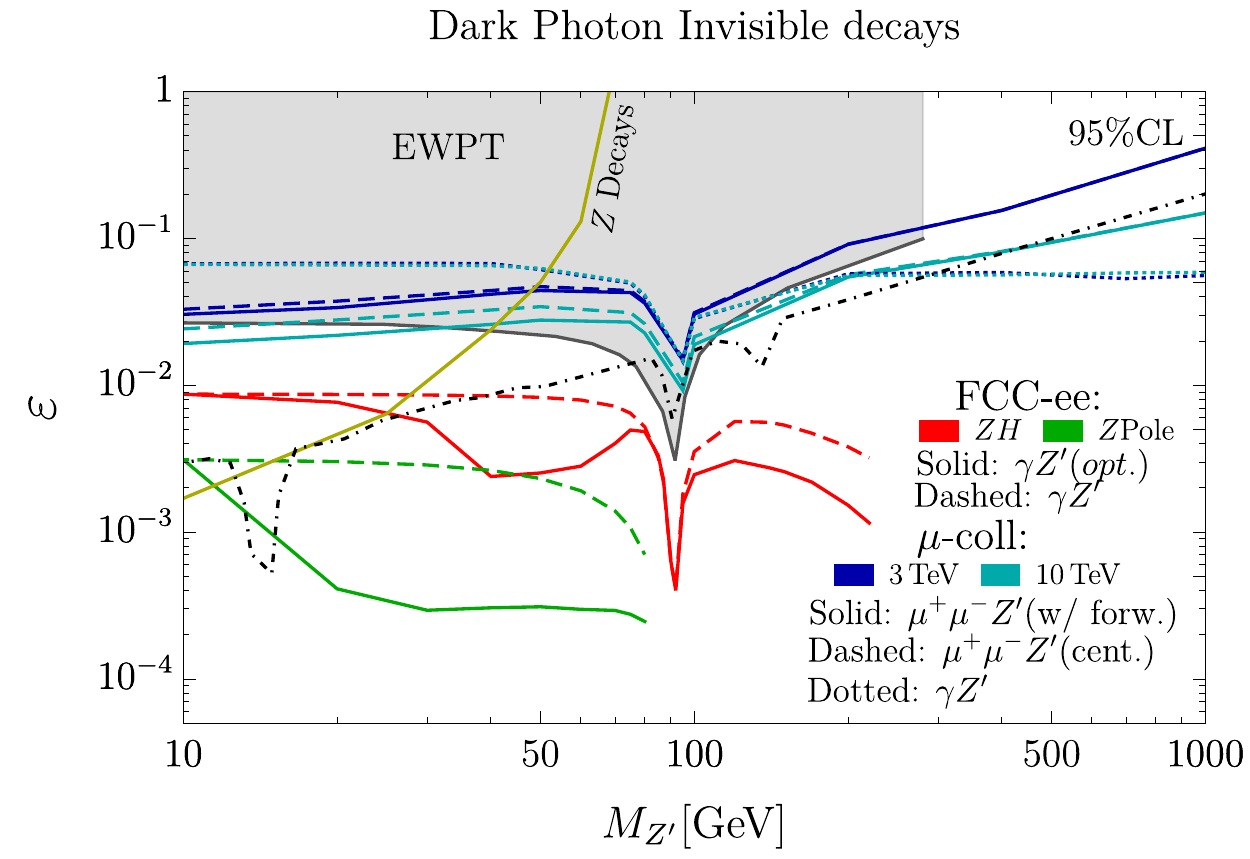}
\end{minipage}
\hspace{0.05\linewidth}
\begin{minipage}{0.45\linewidth}
    \centering
\includegraphics[width=1\linewidth]{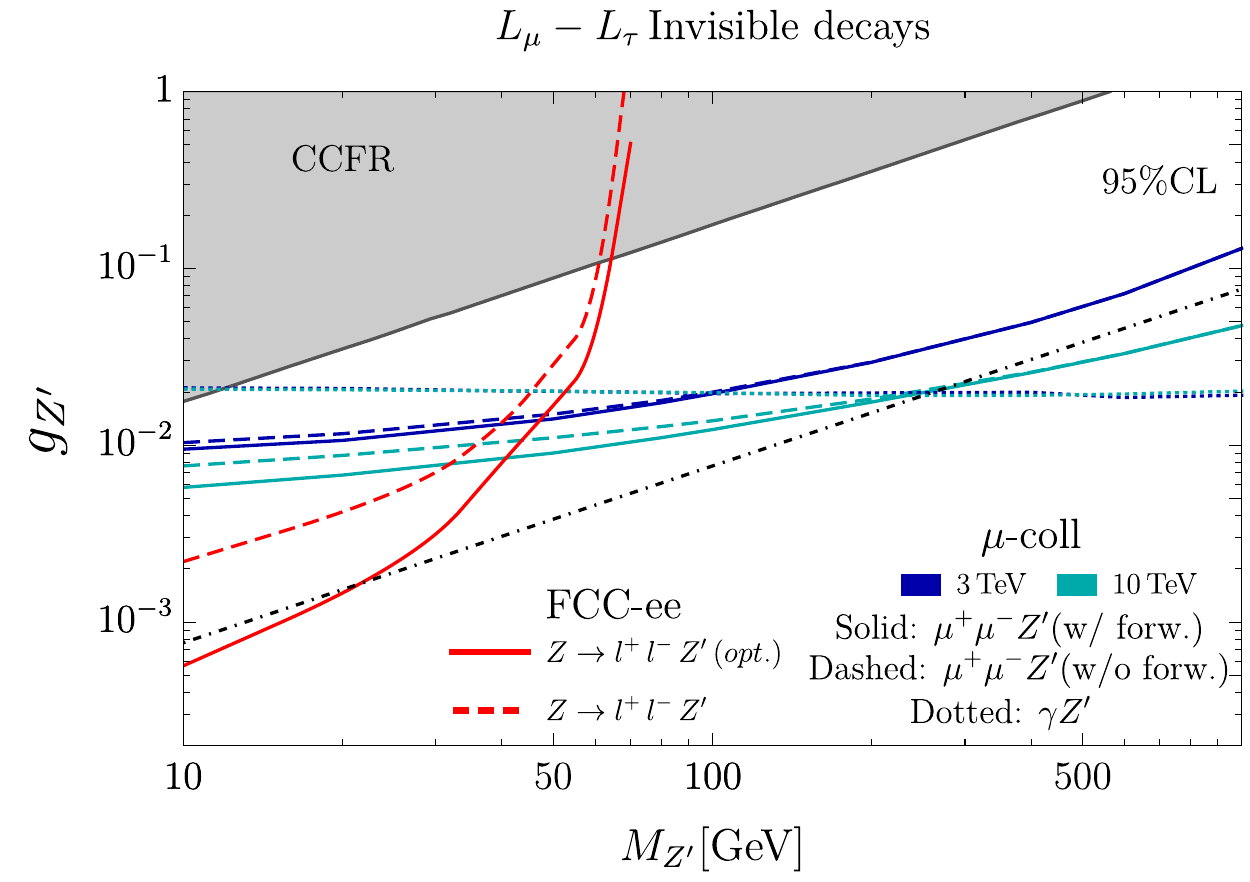}
\end{minipage}
  \caption{Projected sensitivity of future lepton colliders to invisibly decaying $Z'$s. Left: The dark photon portal. Right: The $L_\mu-L_\tau$ portal. The black dot dashed line is the abundance line for fermionic dark matter, $\chi$, with a mass $m_\chi = M_{Z'}/3$, and $\alpha_D =0.1$. It should be noted that in the dark photon case the dark matter must be pseudo-dirac to avoid direct detection constraints \cite{CarrilloGonzalez:2021lxm}. The grey shaded regions are already excluded using limits from EWPT~\cite{Curtin:2014cca} (left) and CCFR~\cite{Altmannshofer:2014pba} (right).}
    \label{fig:Inv_Bound}
\end{figure}

The results of all of our sensitivity projections for the invisibly decaying case may be found in Figure~\ref{fig:Inv_Bound}.
The dark photon portal is shown in the left panel, and $L_{\mu} - L_{\tau}$ is shown in the right panel.

Starting with the dark photon portal it is clear that the FCC-ee can probe a large new portion of parameter space. The dominant production channel, in both the $Z-$pole and the $ZH$ runs, is associated production with a $\gamma$ (see left panel of Fig.~\ref{fig:MuoCZpProd}). As shown in the figure, the derived bounds have a strong dependence on the experimental sensitivity to the photon energy used to reconstruct the mass of the $Z'$. This is particularly relevant for light masses, where a small error in the photon energy can result in a large error in the reconstructed $Z'$ mass, as shown by Eq.~\eqref{eq:monop_smear}. Furthermore, for light masses (below $20$ GeV), where the reconstruction is particularly inefficient, considering exotic $Z$ decays increases the sensitivity. It is worth emphasizing that the FCC-ee can reach the dark matter thermal relic target for most of the relevant mass range, as detailed in the main text. 

The sensitivity of the muon collider, on the contrary, is strongly limited by luminosity. In fact the projected reach for lower masses is well below the FCC-ee, and comparable to present constraints from LEP \cite{Fox:2011fx}. However, the larger CoM energy gives the potential to extend the LEP sensitivity to larger $Z'$ masses by associated production of the $Z'$. The $Z'$-bremsstrahlung from $\mu^+ \mu^-$ scattering (see central panel of Fig.~\ref{fig:MuoCZpProd}) mildly improves the sensitivity with respect to the associated production for masses below $M_Z$. We find that the forward $\mu$ detector provides only minimal improvements in sensitivity, primarily due to the inability to reconstruct the MIM, and the larger backgrounds in the forward region. Finally, we should mention that the FCC-ee is also capable of testing dark photons indirectly through improved determinations of the electroweak precision observables. A back-of-the-envelope estimate, based on the projected sensitivity \cite{deBlas:2022ofj}, suggests an improvement in \(\epsilon\) of roughly an order of magnitude compared to current electroweak precision tests~\cite{Curtin:2014cca}.

Turning to the $L_{\mu} - L_{\tau}$ portal, the $\mu$-collider is clearly the perfect machine for invisible $Z'$s directly coupled to muons and not electrons. This is shown on the right-panel of Fig.~\ref{fig:Inv_Bound} where the $\mu$-collider has the potential to probe a large new portion of parameter space. Associated production is the most promising channel for larger masses ($M_{Z'}\gtrsim 100$ GeV). Similar to the dark photon case, lighter masses can be better tested by the $Z'$-bremsstrahlung from $\mu \mu$ scattering. We again find that forward detectors give only a minimal improvement in sensitivity.

The FCC-ee is an efficient probe for this model for masses below $ M_{Z}$. At the $Z$-pole the copious production of $Z$s allows the production of the $L_{\mu} - L_{\tau}$ gauge boson via exotic $Z$ decays (see right panel of Fig.~\ref{fig:MuoCZpProd}). Our projections show that the region probed by future colliders sits just above the thermal reference line for fermionic dark matter, $\chi$, with a mass $m_\chi=M_{Z'}/3$. It should be noted that for a smaller ratio of $m_\chi/M_{Z'}$ the abundance line moves up into the region of sensitivity. In summary both colliders can surpass the current leading sensitivity (provided by neutrino trident production at the CCFR experiment \cite{Altmannshofer:2014pba}) achieving more than an order of magnitude improvement across a wide range of masses.

\subsection{Decays to the Standard Model}

\begin{figure}
\begin{minipage}{0.45\linewidth}
    \centering
    \includegraphics[width=1\linewidth]{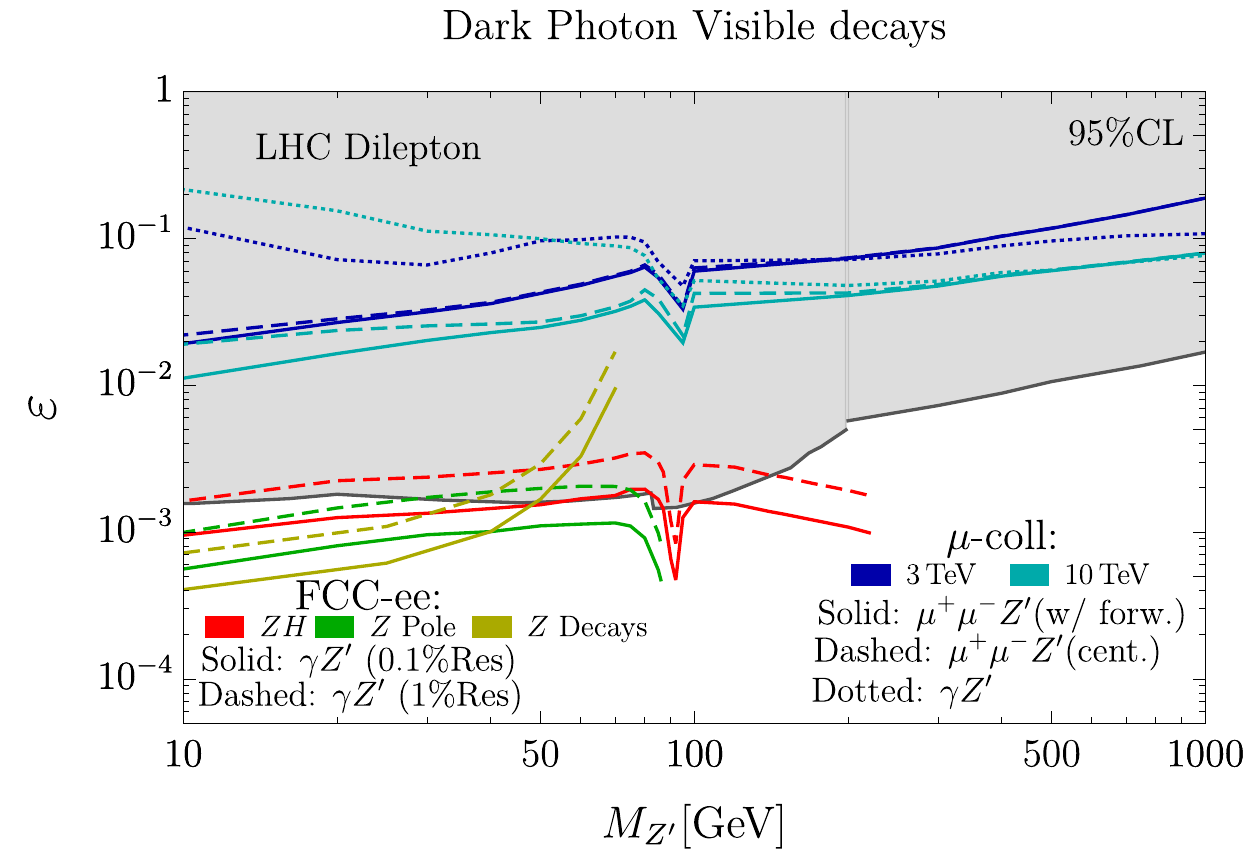}
\end{minipage}
\hspace{0.05\linewidth}
\begin{minipage}{0.45\linewidth}
    \centering
\includegraphics[width=1\linewidth]{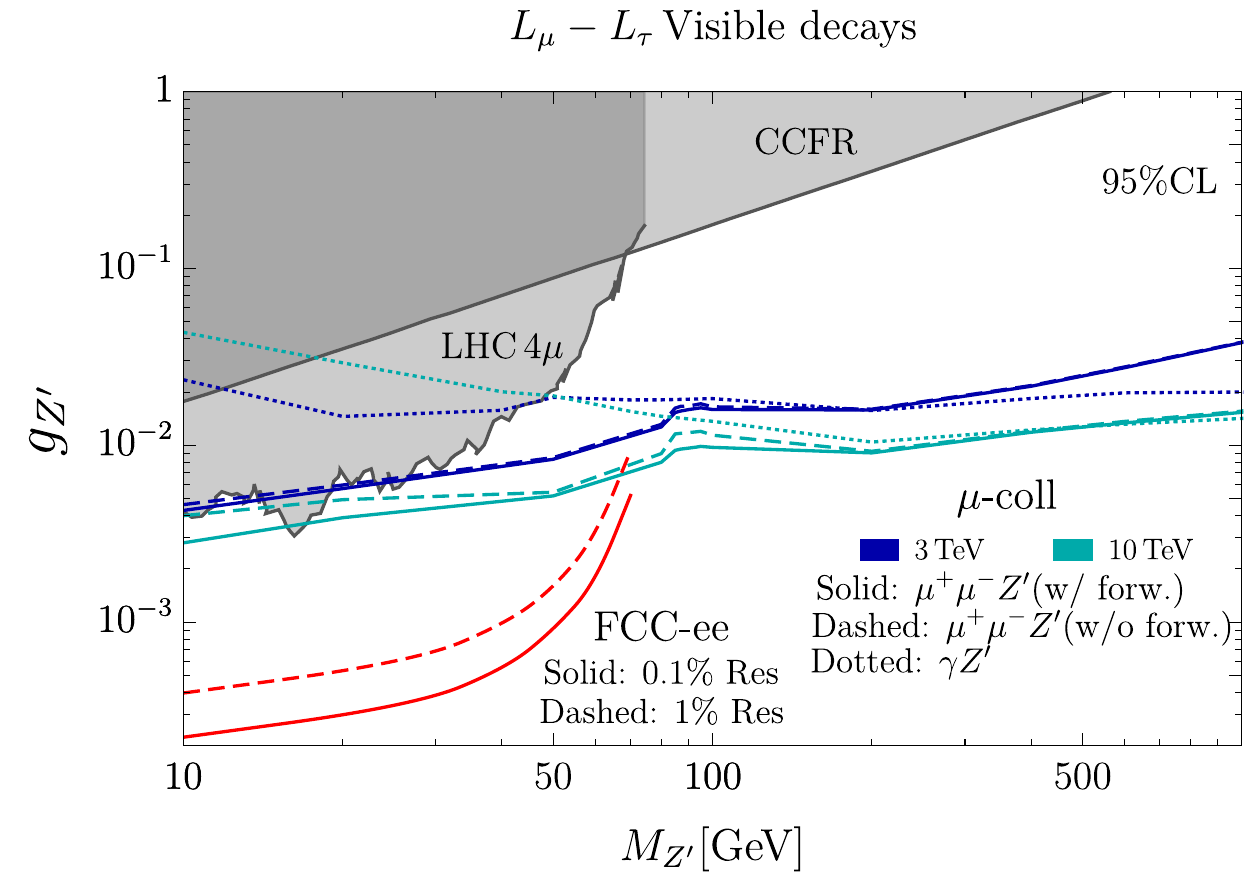}
\end{minipage}
  \caption{Projected sensitivity of future lepton colliders to visibly decaying $Z'$s. Left: dark photon. Right: $L_\mu-L_\tau$. The grey shaded regions are already excluded using limits from LHC dimuon searches~\cite{CMS:2024zqs,CMS:2021ctt} (left) and $4\mu$ searches at LHC~\cite{ATLAS:2023vxg, CMS:2018yxg} (right). }
    \label{fig:Vis_Bound}
\end{figure}

Sensitivity projections for the visibly decaying $Z'$ are shown in Fig.~\ref{fig:Vis_Bound} for the dark-photon (left-panel) and $L_{\mu}-L_{\tau}$ (right-panel).

For the dark photon we observe that, as anticipated in the introduction, only mild improvements are expected from future lepton colliders. Indeed, lepton colliders have to compete with the extremely large number of events produced at the LHC, especially when the \( Z' \) couples to quarks as the visible leptonic final states can often be identified, even at a hadron collider. We find that the \( \mu \)-collider is not competitive with the LHC below a TeV, while the FCC-ee might improve the LHC sensitivity by a factor of $\sim 2$ at the lowest masses.

By contrast, a large new parameter space can be extensively explored for \(L_{\mu} - L_{\tau}\), as shown in the right panel of Fig.~\ref{fig:Vis_Bound}. Starting with the FCC-ee, the copious production of \(Z\) bosons allows for detailed tests of its exotic decays. In particular the \(4\mu\) decays mediated by \(L_{\mu} - L_{\tau}\) result in improved sensitivity by more than an order of magnitude compared to current LHC searches \cite{CMS:2021far,ATLAS:2020uiq} for masses below \(M_Z\). In addition to this the sensitivity
is well above expected HL-LHC projections \cite{Fei:2024qtu}.

The \(\mu\)-collider is more effective for masses above \(M_Z\), as it can probe couplings of order \(10^{-2}\) across a wide range of masses. Furthermore, as we already noticed in the previous scenarios, associated production is favored for masses $M_{Z'} \gtrsim 100 $GeV, while smaller masses can be better tested by the $Z'$-bremsstrahlung in $\mu \mu$ scattering. The CoM energy and the addition of forward $\mu$-detectors only marginally affects the sensitivity reach.\footnote{Additional results with different choices for the forward collider angular resolution can be found in Fig.~\ref{fig:dp_inv_30tev}.}

\section*{Acknowledgments}
We thank R. Capdevilla, M. Ruhdorfer, A. Wulzer for useful discussions. We are grateful to Zhen Liu for reading the manuscript and providing valuable feedback.

S.~A., E.~B. and L.~R. are supported by NSF grant PHY-2210361 and the Maryland Center for Fundamental Physics. GMT is supported in part by the National Science Foundation under Grant Number PHY-2412828.

\appendix

\section{Additional Muon collider projections}\label{App:AddResults}

\begin{figure}[t]
    \centering
    \begin{minipage}[t]{0.48\linewidth}
        \centering
        \includegraphics[width=\linewidth]{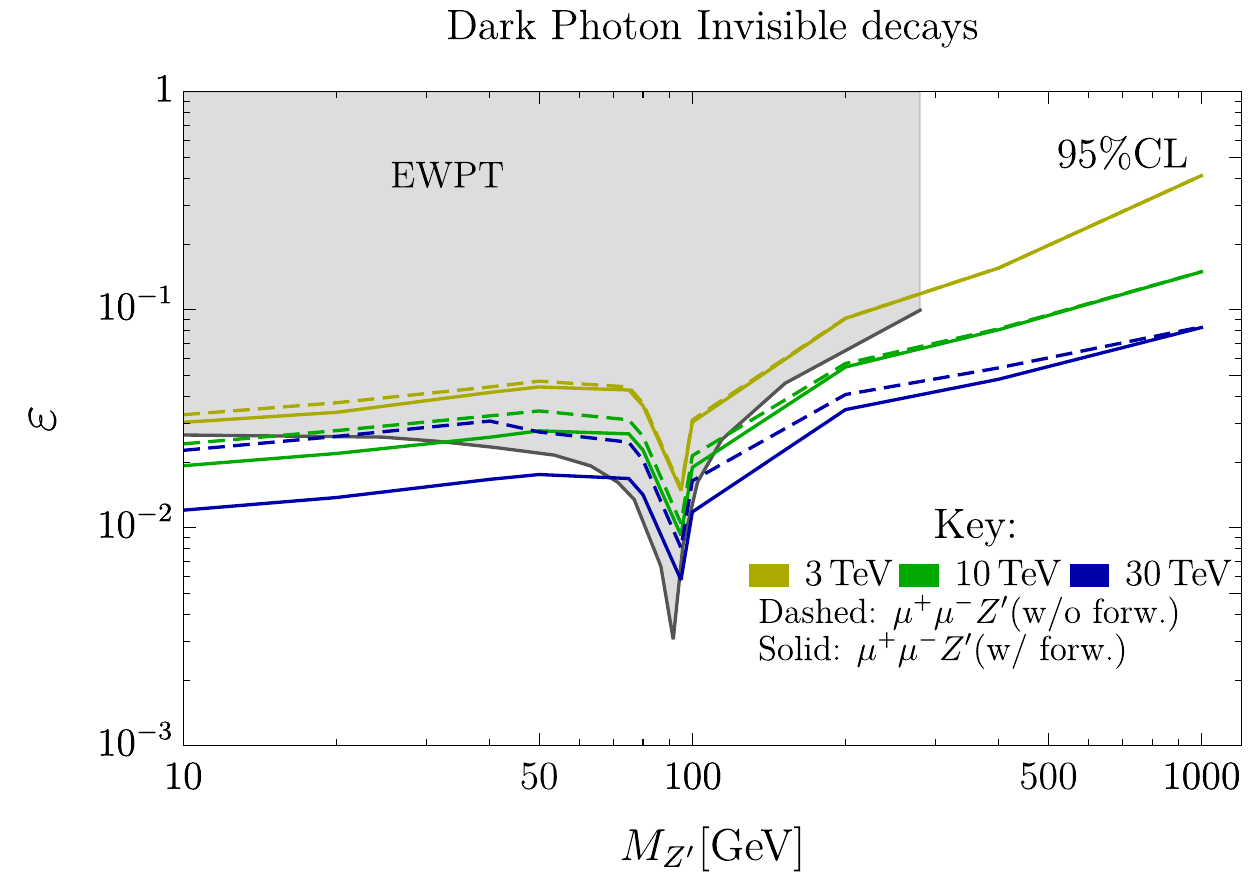}
    \end{minipage}
    \hfill
    \begin{minipage}[t]{0.48\linewidth}
        \centering
        \includegraphics[width=\linewidth]{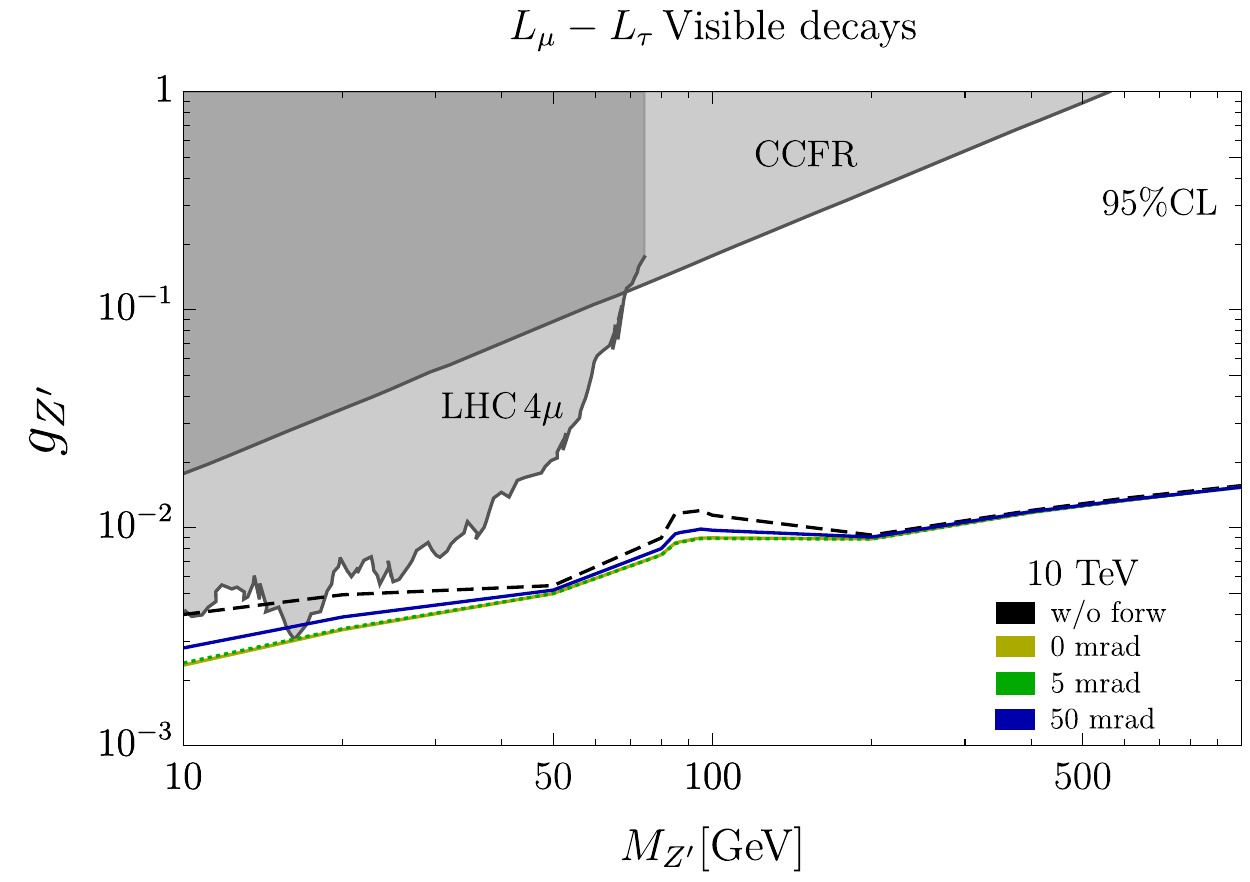}
    \end{minipage}
    \caption{Projected $\mu$-collider sensitivity for $Z'$ bremsstrahlung (middle panel of Fig.~\ref{fig:MuoCZpProd}). Left panel: reach on invisibly decaying dark-photons for different collider energies and assuming (solid) or not (dashed) forward muon detectors. Right-panel: reach on visibly decaying \(L_\mu - L_\tau\) for 10 TeV CoM energies and different angular resolution for the forward $\mu$-detectors.}
    \label{fig:dp_inv_30tev}
\end{figure}

Here we show additional sensitivity projections for the $Z'$-bremsstrahlung (middle panel of Fig.~\ref{fig:MuoCZpProd}) at the $\mu$-collider.

In the left panel of Fig.~\ref{fig:dp_inv_30tev}, we show the 30 TeV CoM energy projections for invisibly decaying dark photons. We observe that the higher energy enhances the production of \( Z' \) bosons, thereby increasing the sensitivity. Additionally, the larger CoM energy favors the radiation of relatively low-\( p_T \) \( Z' \) bosons, which leads to a greater expected improvement from the installment of the forward detectors.

In the right panel of Fig.~\ref{fig:dp_inv_30tev}, we show the impact of varying the angular resolution of the forward \( \mu \)-detector on the sensitivity to the visible \( L_{\mu}-L_{\tau} \). The plot shows how this minimally impacts the sensitivity.

\bibliographystyle{JHEP}
\bibliography{main}
\end{document}